\documentclass{book}    
\usepackage{piers}  
\begin{document}

\title{Analysis of mathematical techniques for the calculation of the electrostatic field in a dielectric-loaded waveguide}
\maketitle

\author      {A. Berenguer}
\affiliation {Dpto. de Ingenier\'ia de Comunicaciones. Univ. Miguel Hern\'andez de Elche}
\address     {}
\city        {Elche (Alicante)}
\postalcode  {03203}
\country     {Spain}
\phone       {}    
\fax         {}    
\email       {aberenguer@umh.es}  
\misc        { }  
\nomakeauthor

\author      {A. Coves}
\affiliation {Dpto. de Ingenier\'ia de Comunicaciones. Univ. Miguel Hern\'andez de Elche}
\address     {}
\city        {Elche (Alicante)}
\postalcode  {03203}
\country     {Spain}
\phone       {}    
\fax         {}    
\email       {angela.coves@umh.es}  
\misc        { }  
\nomakeauthor

\author      {E. Bronchalo}
\affiliation {Dpto. de Ingenier\'ia de Comunicaciones. Univ. Miguel Hern\'andez de Elche}
\address     {}
\city        {Elche (Alicante)}
\postalcode  {03203}
\country     {Spain}
\phone       {}    
\fax         {}    
\email       {ebronchalo@umh.es}  
\misc        { }  
\nomakeauthor

\author      {F. Mesa}
\affiliation {Dpto. de F\'isica Aplicada I. Univ. de Sevilla}
\address     {}
\city        {Sevilla}
\postalcode  {41012}
\country     {Spain}
\phone       {}    
\fax         {}    
\email       {mesa@us.es}  
\misc        { }  
\nomakeauthor

\author      {B. Gimeno}
\affiliation {Dpto. de F\'isica Aplicada y Electromagnetismo-Inst. de Ciencia de Materiales. Univ. de Valencia}
\address     {}
\city        {Valencia}
\postalcode  {46100}
\country     {Spain}
\phone       {}    
\fax         {}    
\email       {benito.gimeno@uv.es}  
\misc        { }  
\nomakeauthor

\begin{authors}

{\bf A. Berenguer}$^{1}$, {\bf A. Coves}$^{1}$, {\bf E. Bronchalo}$^{1}$ and {\bf F. Mesa}$^{2}$\\
\medskip
$^{1}$Dpto. de Ingenier\'ia de Comunicaciones. Univ. Miguel Hern\'andez de Elche, Spain\\
$^{2}$Dpto. de F\'isica Aplicada I. Univ. de Sevilla, Spain\\

\end{authors}

\begin{paper}

\begin{piersabstract}
The resolution of the Green's function for obtaining the electrostatic potential generated by the charges located in the dielectric layer of a rectangular waveguide requires efficient integration techniques. Due to the characteristics and the oscillatory nature of the function to be integrated, the use of the Filon's method together with a convergence analysis is adequate. However, the application of this numerical integration technique can lead to numerical instabilities in the result. For this reason, in this research work we have presented two different methods to deal with these numerical errors of integration: SSA/MSSA and GPR. In this way it is possible to clean the electrostatic potential avoiding subsequent errors in the calculation of the electrostatic field. 
\end{piersabstract}

\psection{Introduction}
The calculation of the electrostatic field, Edc, in a dielectric-loaded waveguide due to an arbitrary charge distribution on the dielectric layer is a problem of great interest in the space industry, because of the lack of rigorous studies about the multipactor effect appearing in dielectric loaded waveguide-based microwave devices in satellite on-board equipment. When dealing with a partially dielectric-loaded rectangular waveguide, the electrons emitted by the dielectric surface charge the dielectric material positively, whereas the electrons absorbed by the dielectric layer charge it negatively. This charge gives rise to an electrostatic field which has to be taken into account in order to obtain an accurate trajectory of the electrons in the structure. While many works have studied the electrostatic field appearing on RF dielectric windows \cite{Ang,Neuber,Kishek,Valfells,Valfells2,Yla,Anderson,Michizono}, less attention has received the analgous problem in dielectric loaded waveguides \cite{Torregrosa,Coves,Torregrosa2}. Although the problem of obtaining the electrostatic field originated by an arbitrary electron charge distribution has been solved in many electromagnetism books \cite{Griffiths,Jackson,Bladel}, the high computational complexity of this problem requires the use of different mathematical techniques to minimize computation time. In addition, there may be critical points, e.g. close to the walls or the dielectric layer, in the structure in which the solution does not converge properly. The convergence of the solution to the problem of calculating the electrostatic potential necessary for the calculation of the electrostatic field sometimes presents numerical instabilities generating false peaks. In this work, several numerical algorithms have been analysed and compared for the detection of these outliers in the electrostatic potential. These include, among others, Spectrum Analysis Models (e.g. SSA/MSSA) or Non-parametric Bayessian Models (e.g. GPRs).

The paper is organized as follows: Section \ref{s:Theory} describes the theory and fundamental principles underlying the problem under investigation. In Section \ref{s:Results}, the results obtained for a particular rectangular waveguide is shown, including some examples of the application of SSA/MSSA and GPR methods. A few concluding remarks are made in Section \ref{s:Conclusion}.

\psection{Theory}\label{s:Theory}

\psubsection{Electrostatic field in a dielectric-loaded waveguide due to an arbitrary charge distribution on the dielectric layer}
In Fig. \ref{esquema} it is shown the section of the waveguide under study, consisting on a partially dielectric-loaded rectangular waveguide, whose dielectric layer has relative permittivity $\epsilon_r$ and thickness $h$.
\begin{figure}[h!]
	\centering
	\includegraphics[width=0.45\textwidth]{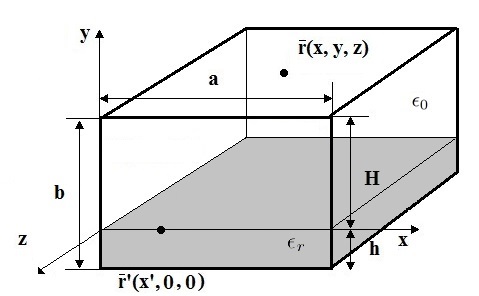}
	\caption{Geometry and dimensions of the problem under investigation.}\label{esquema}
\end{figure}
The aim is to compute the electrostatic field at the observation point $\mathbf{r} = (x,y,z)$, which is assumed to be located in the air region of a waveguide with translational symmetry along the longitudinal direction $z$, due to a point charge on the dielectric layer at $\mathbf{r'}=(x',0,0)$. 

The electrostatic field $\textbf{E}_\text{DC}$ generated by the charges on the dielectric can be obtained as,
\begin{equation} \label{Edc}
	\textbf{E}_\text{DC}(x,y,z) = -\nabla\phi(x,y,z) 
\end{equation}
where $\phi(x,y,z)$ is the potential inside the waveguide. Using superposition, this potential due to the set of charges $Q_{i}$ on the dielectric surface can be obtained by adding the individual contribution of each charge,
\begin{equation} \label{superposition}
	\phi (x,y,z) = \sum_{i}{G(x-x_{i}',y,\left|z-z_{i}'\right|)Q_{i}(x_{i}',0,z_{i}')} 
\end{equation}
where $G(x,y,z)$ is the electrostatic potential due to a unit point charge, that is, the Green's function for this problem. Due to the geometric characteristics and the linear nature of the problem under consideration  it is straightforward to demonstrate, as explained in \cite{berenguer16,berenguer19}, that the Green's function in the spatial domain $G$ can be obtained in the air region $y \geq 0$ as,
\begin{equation} \label{G}
	G(x,x',y,z) = \frac{2}{\epsilon_0 \pi a} \sum^{\infty}_{n=1}\sin(k_{xn}x)\sin(k_{xn}x') 
	\times\int^{\infty}_{0}\frac{\sinh[k_t(H-y)]\cos(k_zz)}{k_t[\epsilon_r
	\coth(k_th)+\coth(k_td)]\sinh(k_td)}\,\mathrm{d} k_z\;.
\end{equation}
In \eqref{G}, if the point charge is placed at $z'\neq 0$, $z$ must be replaced by $(z-z')$. Here it is worth noting that very efficient numerical summation and integration techniques have to be employed to compute the Green's function with sufficient accuracy and tolerable CPU times. Because of the rapid oscillation of the integrand for large values of $z$, Filon's integration method is chosen since it is desirable for integrals that has the form \cite{Hildebrand},
\begin{equation} \label{Filon}
	\int^{b}_{a}f(x)\cos(kx)\mathrm{d}x
\end{equation}

Using superposition, the potential due to an arbitrary charge distribution on a dielectric layer is obtained by adding the individual contribution of each point charge, 
\begin{equation} \label{superposition}
\phi (x,y,z) = \int{G(x-x',y,z)\rho(x')\mathrm{d}x'}
\end{equation}

Once the Green's function has been calculated, the $\textbf{E}_\text{DC}$ field is obtained by using (\ref{superposition}) and calculating numerical differentiation of (\ref{Edc}) by means of the central difference technique.

\psubsection{Spectrum Analysis Models: SSA and MSSA}
The Singular Spectrum Analysis (SSA), is a relatively novel but powerful technique in data analysis, which has been developed and applied to many practical problems across different fields. The SSA technique is a decomposition-based approach and its usefulness lies in extracting information from the (auto)covariance structure of the data. In the original formulation of SSA it was assumed that the data under analysis has a deterministic component with noise superimposed and that the deterministic component can be successfully extracted from the noise. This formulation is not, of course, confined to SSA; the decomposition-based approach to data analysis is very old. What SSA brings into the picture, that identify it as a novel method, is that it accounts for the (auto)covariance structure of the data without imposing a parametric model for it. It is thus a data adaptive, non-parametric method based on embedding a data in a vector space, and from a practical perspective, a model-free approach for data analysis. The SSA method proceeds by diagonalizing the lag-covariance matrix to obtain spectral information on the data. The data can then be analyzed as a sum of simpler, elementary series which correspond to different sub groups of eigentriples (each eigentriple is composed of an eigenvalue and its associated left and right eigenvectors) of the lag-covariance matrix.
\psubsubsection{Step 1}
The first step, called embedding, is to transform a one-dimensional data $\{x_1, ..., x_n\}$ into a multidimensional trajectory matrix of lagged vectors $\textbf{\textit{X}} = [\textbf{x}_1, ..., \textbf{x}_k]$ where $k = n - m + 1$ and each lagged vector is defined as $\textbf{x}_i = (x_i, ..., x_{i+m-1})^T$ for $i = 1, ..., k$. Each of these vectors corresponds to a partial view of the original data, seen through a window of length $m$. Choosing the lag window size $m$ is a matter of balancing the retrieval of information on the structure of the underlying data and the degree of statistical confidence in the results. The trajectory matrix $\textbf{\textit{X}}$ is a rectangular Hankel matrix of the form,
\begin{equation}
	\textbf{\textit{X}} = \begin{bmatrix} x_1 & x_2 & \cdots & x_k \\ x_2 & x_3 & \cdots & x_{k+1} \\ \vdots & \vdots & \vdots & \vdots \\ x_m & x_{m+1} & \cdots & x_{n} \end{bmatrix} 
\end{equation}

\psubsubsection{Step 2}

The second step consists on the Singular Value Decomposition (SVD) of the trajectory matrix $\textbf{\textit{X}}$ as,
\begin{equation}
	\textbf{\textit{X}} = \textbf{\textit{U}} \textbf{\textit{$\Sigma$}} \textbf{\textit{V}}^T
\end{equation}
where \textbf{\textit{U}} is an orthogonal matrix of size $m \times m$, \textbf{\textit{$\Sigma$}} is a rectangular diagonal matrix of
size $m \times k$ and \textbf{\textit{V}} is an orthogonal matrix of size $k \times k$. The elements of \textbf{\textit{$\Sigma$}}, called singular values, are the square roots of the eigenvalues of the covariance matrix $\textbf{\textit{X}}\textbf{\textit{X}}^T$. The rows of
\textbf{\textit{U}} are the eigenvectors of $\textbf{\textit{X}}^T\textbf{\textit{X}}$ and are referred to as the left singular vectors. The columns of $\textbf{\textit{V}}^T$ are the eigenvectors of $\textbf{\textit{X}}\textbf{\textit{X}}^T$. For SSA, the singular values are organized in decreasing order. Then any subset of the $d$ eigentriples $1 \leq d \leq m$, for which the singular values are strictly positive provides the best representation of the $\textbf{\textit{X}}$ matrix as a sum of $\textbf{\textit{X}}_i$ matrices for $i = 1, ..., d$.

\psubsubsection{Step 3}
The third step involves the partitioning of these $d$ eigentriples into $p$ disjoint sub groups and summing them within each group, such that
represents a component series described by distinct subsets of eigentriples.

\psubsubsection{Step 4}
The last step of the SSA algorithm, known as \textit{diagonal averaging}, aims at transforming the component matrices $\textbf{\textit{X}}_i$ into Hankel matrices, which then become the trajectory matrices of the underlying data, in such a way that the original data can be reconstructed as a sum of these components. The entire procedure aims at defining in some optimal way what those components are.

For multivariate data, the Multichannel Singular Spectrum Analysis (MSSA) gap filling algorithm takes advantage of both spatial (cross multiple data) and (auto)correlation.

\psubsection{Non-parametric Bayessian Models: Gaussian Process Regressions (GPRs)}
Gaussian Processes (GPs) provide an alternative approach to regression problems. The GP approach is a non-parametric approach, in a way that it finds a distribution over the possible functions $f(x)$ that are consistent with the observed data,
\begin{equation}
	y = f(x) + \epsilon
\end{equation}
As with all Bayesian methods it, begins with a prior distribution and updates this as data points are observed, producing the posterior distribution over functions. A GP defines a prior over functions, which can be converted into a posterior over functions once we have
seen some data. Although it might seem difficult to represent a distrubtion over a function, it turns out that we only need to be able to define a distribution over the function's values at a finite, but arbitrary, set of points $\{x_1, ..., x_n\}$. A GP assumes that $p(f(x_1)), ..., p(f(x_n))$ is jointly Gaussian, with some mean $\mu(x)$ and covariance matrix $\Sigma(x, x')$ given by $\Sigma_{ij} = k(x_i, x_j)$, where $k$ is a positive definite kernel function,
\begin{equation}
	\begin{bmatrix} f(x_1) \\ \vdots \\ f(x_n) \end{bmatrix} 
	\sim \mathcal{N} \left(\begin{bmatrix} \mu(x_1) \\ \vdots \\ \mu(x_n) \end{bmatrix}, \begin{bmatrix} k(x_1, x_1) & \cdots & k(x_1, x_n) \\ \vdots & \vdots & \vdots \\ k(x_n, x_1) & \cdots & k(x_n, x_n) \end{bmatrix} \right)
\end{equation}
The key idea is that if $x_i$ and $x_j$ are deemed by the kernel to be similar, then we expect the output of the function at those points to be similar, too. The mathematical crux of GPs is the multivariate Gaussian distribution. The covariance matrix, along with a mean function to output the expected value of $f(x)$ defines the Gaussian Process.

Since the key assumption in GP modelling is that our data can be represented as a sample from a multivariate Gaussian distribution, we have that,
\begin{equation}
	\begin{bmatrix} f \\  f_* \end{bmatrix} 
	\sim \mathcal{N} \left(\begin{bmatrix} \mu \\ \mu_* \end{bmatrix}, \begin{bmatrix} K & K_*^T \\ K_* & K_{**} \end{bmatrix} \right)
\end{equation}
where $K$ is the matrix we get by applying the kernel function to our observed $x$ values (i.e. training data), $K_*$ gets us the similarity of the observed $x$ values to the values whose output we're trying to estimate (i.e. test data) and $K_{**}$ gives the similarity of the estimated values to each other. $T$ indicates matrix transposition. 

We are interested in the conditional probability $p(f_*|f)$: "given the data, how likely is a certain prediction for $f_*$?". The probability follows a Gaussian distribution,
\begin{equation}
	f_*|f \sim \mathcal{N}(K_*K^{-1}f, K_{**}-K_*K^{-1}K_*^T)
\end{equation}
Our best estimate for $f_*$ is the mean of this distribution,
\begin{equation}
	\bar{f}_* = K_*K^{-1}f
\end{equation}
and the uncertainty in our estimate is captured in its variance,
\begin{equation}
	\mathrm{var}(f_*) = K_{**}-K_*K^{-1}K_*^T
\end{equation}

\psection{Numerical results and discussion}\label{s:Results}
This section shows the results obtained for a particular rectangular waveguide. In this case, the following parameters for the geometry and materials are considered: $a=19.05$ mm, $H=0.375$ mm, $h=0.25$ mm, $z=1$ mm and $\epsilon_r=2.25$.

The shape of the spectral Green's function, eq.(\ref{G}), with respect to the integration variable $k_z$ has been analyzed into detail. Fig. \ref{integral} shows the shape of the function to be integrated for the first term of the summation, $n = 1$, and a distance $y=0.1$ mm from the surface of the dielectric layer.
\begin{figure}[h]
	\centering
	\includegraphics[width=0.9\textwidth]{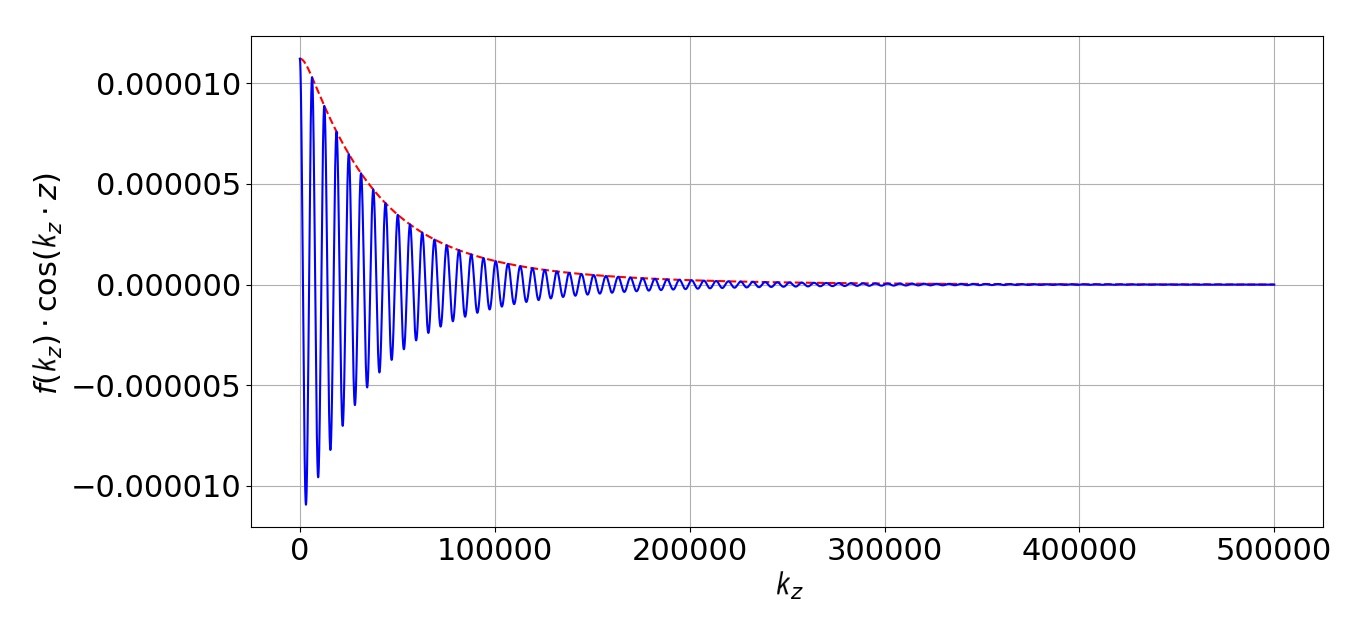}
	\caption{The integrated function for the first term of the summation, $n = 1$, and a distance $y=0.1$ mm.}\label{integral}
\end{figure}
 
 In order to achive accurate results, it is relevant to carry out a convergence study on the integrand. In terms of the rate of convergence, the worst scenarios are for the cases of low $y$ values and high $n$ values. $f(k_z)$ for the case of $y=0.1$ mm and $n=1$ is plotted in Fig. \ref{integral_log}. As it is shown, $k_z\geq4\times10^4$ has to be considered to reach convergence. The asymptotic behavior of the integrand, determined by the term $e^{-k_ty}$, allows us to establish a condition to stop the computation when the convergence is reached. It involves calculating the relative value of the i-th summand of the integral with respect to the accumulated value of the integral until this iteration. If this relative value is less than a particular convergence tolerance, the computation of the integral is stopped.
\begin{figure}[h]
	\centering
	\includegraphics[width=0.9\textwidth]{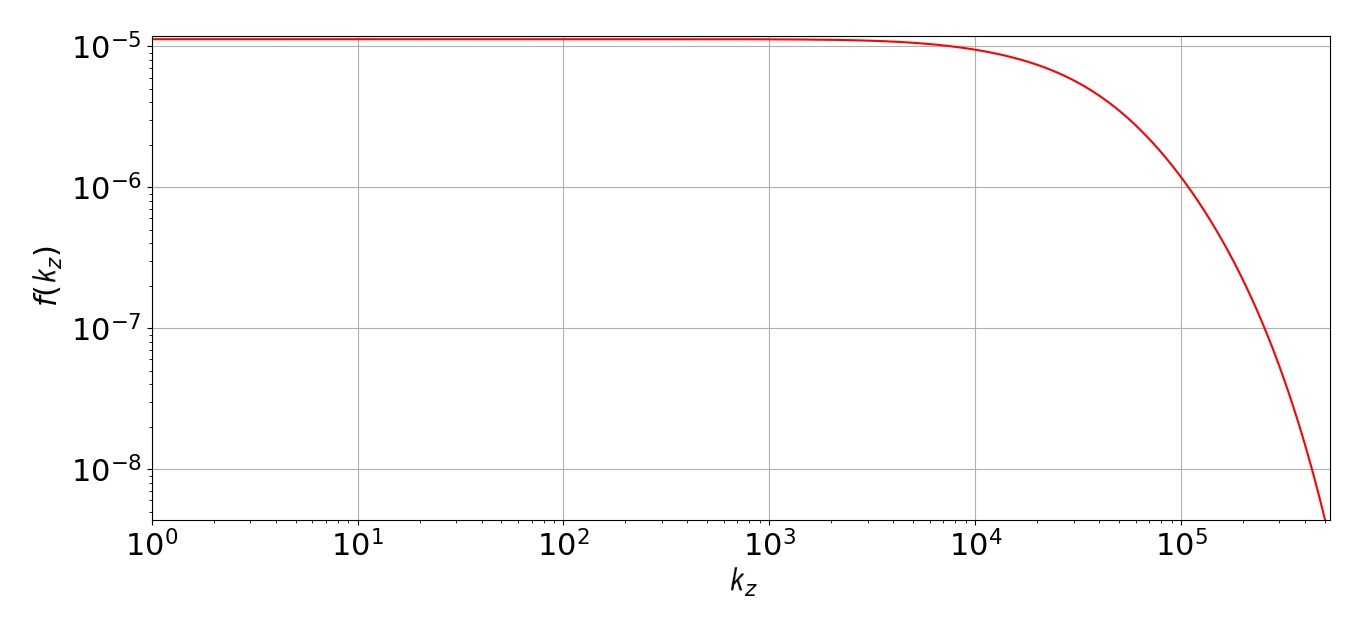}
	\caption{$f(k_z)$ for the case $a=19.05$ mm, $H=0.375$ mm, $h=0.25$ mm, $y=0.1$ mm, $z=1$ mm, $n=1$ and $\epsilon_r=2.25$.}\label{integral_log}
\end{figure}

As shown in Fig. \ref{potential}, false peaks are observed in the the Green's function calculated by using Filon's numerical integration. To fix these errors and remove them, the SSA/MSSA technique is used. To do this, a lag window of size $m = 4$ is considered for calculating the trajectory matrix of the electrostatic potential. The reason for choosing this size is that the objective is to minimize or eliminate false peaks when possible, impacting the correct values of the potential as little as possible. Choosing a higher value of the window lag $m$ would lead to reduce these undesirable effects in the electrostatic potential, however it would impact the real value of the centered peak.

In spite of obtaining accurate results with SSA/MSSA technique, a second method of data cleaning is used to compare the results. In this case the GPR method is applied to the potential data. In this case, the results for the electrostatic potential improve slightly the ones obtained from the SSA/MSSA method.  

\begin{figure}[h]
	\centering
	\includegraphics[width=0.9\textwidth]{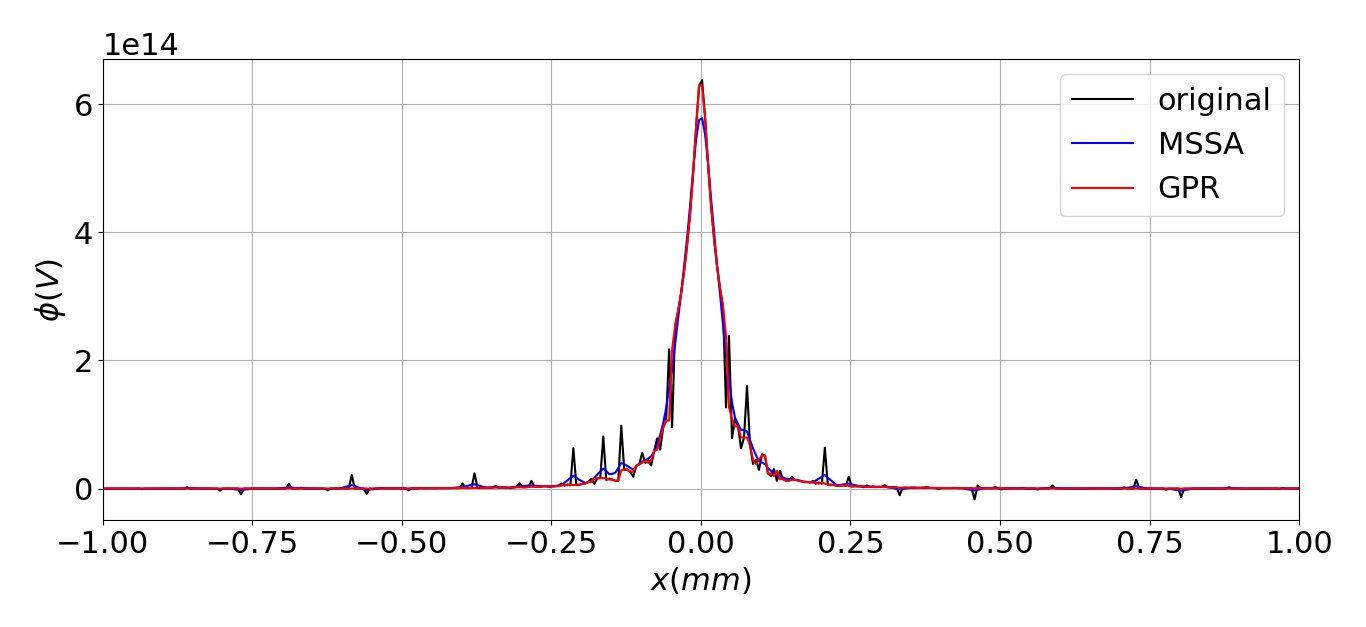}
	\caption{Comparasion of SSA/MSSA and GPR methods for data cleaning process of the electrostatic potential.}\label{potential}
\end{figure}

\psection{Conclusion}\label{s:Conclusion}
In this work, two different numerical methods, SSA/MSSA and GPR, has been shown for dealing with numerical errors when calculating the Green's function. Although Filon's method is suitable for integrating oscillating functions, as is the case of the problem under investigation, for certain points, false peaks in the potential are obtained which need to be corrected to avoid their propagation in the calculation of the electrical field generated by a charge distribution on the surface of the dielectric layer. The GPR behaves slightly better than the MSSA/SSA method in the examples analyzed in this work. However, the results obtained with both of them are accurate enough to solve this issue. 

\ack
This work was supported by the Agencia Estatal de Investigaci{\'o}n (AEI) and by the Uni{\'o}n Europea through the Fondo Europeo de Desarrollo Regional - FEDER - “Una manera de hacer Europa” (AEI/FEDER, UE), under the Research Project TEC2016-75934-C4-2-R.

\end{paper}

\end{document}